\documentclass[12pt]{article}
\begin{document}
\begin{titlepage}
\begin{flushright}
hep-th/0204009 \\
YITP-02-21\\
TIT-HEP-476\\
April, 2002
\end{flushright}
\vspace{0.5cm}
\begin{center}
{\Large \bf 
Hybrid Superstrings on Singular \\Calabi-Yau Fourfolds
}
\lineskip .75em
\vskip2.5cm
{\large Katsushi Ito${}^{1}$} and {\large Hiroshi Kunitomo${}^{2}$}
\vskip 1.5em
${}^{1}${\large\it Department of Physics\\
Tokyo Institute of Technology,Tokyo, 152-8551, Japan}  
\vskip 1.5em
${}^{2}${\large\it Yukawa Institute for Theoretical Physics\\
Kyoto University, Kyoto 606-8502, Japan}
\vskip 3.5em
\end{center}
\begin{abstract}
Two-dimensional hybrid superstring on singular Calabi-Yau manifolds
is studied by the field
redefinition of the NSR formalism. The compactification on
singular Calabi-Yau fourfold is described by $N=2$ super-Liouville
theory and $N=2$ Landau-Ginzburg models. We examine
the world-sheet topological $N=4$ superconformal algebra,
which is useful to identify physical states of the theory.
It is shown that the space-time superconformal symmetry
is not compatible with this original topological algebra.
A new model is proposed by modifying the topological
superconformal generators, which is consistent with
space-time $N=2$ superconformal symmetry.

\end{abstract}
\end{titlepage}
\baselineskip=0.7cm
Holography between type II string theory on a singular Calabi-Yau 
manifold and superconformal field theory provides a useful tool for
studying strong coupling physics of supersymmetric gauge theories
\cite{OoVa,GKP,GiKu,EgSu}.
This string theory in the decoupling limit describes a field theory on the 
NS fivebranes wrapped on cycles on Calabi-Yau manifolds.
When a Calabi-Yau manifold becomes singular, the field theory 
becomes an interacting superconformal field theory.
Giveon, Kutasov and Pelc \cite{GKP} studied the type II superstrings on
a singular hypersurface and calculated the mass spectrum in the
Neveu-Schwarz-Ramond (NSR) formalism.

It is an interesting problem to add the 
Ramond-Ramond (RR) flux in this system.
The superstring theory including the RR flux is recognized more
important to understand the phase structure of related supersymmetric
gauge theories\cite{Va}. 
Since the RR vertex operators are represented by spin fields in the 
NSR formalism, it is difficult  to quantize the superstring theories
in this formulation.
Green-Schwarz formalism, on the other hand, includes the RR fields
naturally. But it is difficult to quantize covariantly due to its 
non-linearity.

Hybrid superstrings proposed by Berkovits \cite{Be} and Berkovits and
Vafa \cite{BeVa}
is a useful approach to quantize the superstrings compactified on 
Calabi-Yau manifolds in a manifestly supersymmetric manner.
This is defined by using only mutually local spin fields which are 
obtained by field redefinition of the NSR formalism.
The worldsheet topological $N=4$ superconformal algebra plays an
important role for the construction of physical states.
One can also introduce the RR flux in this formalism\cite{BeVaWi}.

In this paper we will investigate the hybrid superstrings on singular
Calabi-Yau
manifolds. In particular we study the two dimensional hybrid
superstrings
compactified on Calabi-Yau fourfolds. This model is recently studied
\cite{BeGuVa}.
Singular Calabi-Yau manifolds are described by tensor products of the
$N=2$ 
super-Liouville theory and $N=2$ Landau-Ginzburg models.

The space-time sector of this model includes two scalar fields
$(\rho,f)$
in addition to the superspace coordinate fields.  By combining this
space-time
sector and the $N=2$ Liouville theory with background charge
$Q=\sqrt{\frac{2}{k}}$,
we can naturally obtain the free field realization of the affine Lie
superalgebra
$sl(2|1)^{(1)}$ at level $k$. From this superalgebra, we construct a
space-time
$N=2$ superconformal algebra with central charge $c=6kp$ for integer
$p$.
For nonzero $k$, however, we will find generators of this space-time
superconformal
symmetry do not commute with the BRST operators obtained by the field
redefinition of
the NSR superstring. We propose a model defined by a new topological
$N=4$ superconformal
algebra including BRST currents. This topological algebra is consistent
with space-time
$N=2$ superconformal symmetry.

We begin with reviewing the hybrid superstrings on Calabi-Yau
fourfolds\cite{BeGuVa}. 
We consider the NSR formalism of type II superstrings on
$R^{1,1}\times X^{8}$, where $R^{1,1}$ is two-dimensional Minkowski
space-time
and $X^{8}$ is a Calabi-Yau fourfold.
The flat space-time $R^{1,1}$ is described by worldsheet free bosons
 $X^{m}(z)$ and 
real free fermions $\psi^{m}(z)$ ($m=0,1$) 
with operator product expansions (OPEs)
\begin{equation}
 X^{m}(z)X^{n}(w)\sim -\eta^{mn}\ln(z-w),\quad
\psi^{m}(z)\psi^{n}(w)\sim\eta^{mn}/(z-w).
\end{equation}
Here $\eta^{mn}=diag(-1,1)$.
Introduce the light-cone basis $X^{\pm}={1\over\sqrt{2}}(\pm X^{0}+
X^{1})$, $\psi^{\pm}={1\over\sqrt{2}}(\pm \psi^{0}+
\psi^{1})$. 
We have
\begin{equation}
 X^{+}(z)X^{-}(w)\sim -\ln(z-w),\quad
\psi^{+}(z)\psi^{-}(w)\sim {1\over z-w}.
\end{equation}
The energy-momentum tensor $T_{M}$ and $N=1$ worldsheet supercurrent
$G_{M}$ is given by
\begin{eqnarray}
 T_{M}&=&-\partial X^{+}\partial X^{-}
-{1\over2}\left(\psi^{+}\partial\psi^{-}+\psi^{-}\partial\psi^{+}\right),
\nonumber\\
G_{M}&=& i\psi^{+}\partial X^{-}+i \psi^{-}\partial X^{+}
\end{eqnarray}

The Calabi-Yau part is described by $N=2$ superconformal field theory (SCFT)
with central charge $c_{C}=12$ $(\hat{c}_{C}=c_{C}/3=4)$.
Denote $J_{C},G^{\pm}_{C},T^{N=2}_{C}$ for the generators of $N=2$
superconformal algebra with OPEs
\begin{eqnarray}
T^{N=2}_{C}(z)T^{N=2}_{C}(w)
&\sim&{{c_{C}\over2}\over(z-w)^4}+{2 T^{N=2}_{C}(w)\over (z-w)^2}
+{\partial T^{N=2}_{C}(w)\over z-w},\nonumber\\
T^{N=2}_{C}(z)J_{C}(w)&\sim&{J_{C}(w)\over (z-w)^2}
+{\partial J_{C}(w)\over z-w},
\quad
T^{N=2}_{C}(z)G^{\pm}_{C}(w)\sim{{3\over2}G^{\pm}_{C}(w)\over (z-w)^2}
+{\partial G^{\pm}_{C}(w)\over z-w}
\nonumber\\
 J_{C}(z)J_{C}(w)&\sim&{\hat{c}_{C}\over (z-w)^2},
\quad 
J_{C}(z)G^{\pm}_{C}(w)\sim{\pm G^{\pm}_{C}(w)\over z-w}
\nonumber\\
G^{+}_{C}(z)G^{-}_{C}(w)
&\sim& {\hat{c}_{C}\over (z-w)^3}
+{J_{C}\over (z-w)^2}+{T^{N=2}_{C}(w)+{1\over2}J_{C}(w)\over z-w}
\end{eqnarray}
We need fermionic ghosts $(b,c)$ and bosonic ghosts $(\beta,\gamma)$
with conformal weights (2,-1) and $({3\over2},-{1\over2})$,
respectively. They satisfy
\begin{equation}
 b(z)c(w)\sim{1\over z-w},\quad \beta(z)\gamma(w)\sim{-1\over z-w}.
\end{equation}
These fields are bosonized such as
$(b,c)=(e^{i\sigma},e^{-i\sigma})$,
$(\beta,\gamma)=(\partial\xi e^{-\phi},\eta e^{\phi})$ and
$(\eta,\xi)=(e^{i\chi},e^{-i\chi})$ \cite{FMS}.
Here $\sigma$, $\phi$ and $\chi$ are free bosons with OPEs
$\sigma(z)\sigma(w)\sim -\ln(z-w)$, $\phi(z)\phi(w)\sim -\ln(z-w)$
and $\chi(z)\chi(w)\sim -\ln(z-w)$, respectively.
The ghost sector has $N=1$ superconformal symmetry.
The energy-momentum tensor and the supercurrent are 
\begin{eqnarray}
T^{gh}(z)&=& T_{bc}+T_{\beta\gamma}, \nonumber\\
T_{bc}(z)&=& -2 b\partial c-\partial b c, \quad 
T_{\beta\gamma}(z)=-{3\over2} \beta\partial \gamma
-{1\over2}\partial \beta\gamma, \nonumber\\
G^{gh}(z)&=& 2 c\partial\beta+3\partial c \beta+\gamma b.
\end{eqnarray}
We discuss the BRST structure of the superstring theory. The BRST charge
$Q_{BRST}$ is defined by
\begin{equation}
 Q_{BRST}=\int {d z\over2\pi i} J_{BRST}(z), 
\label{eq:brso}
\end{equation}
where $J_{BRST}$
is the BRST current:
\begin{equation}
J_{BRST}=
c(T_{m}+{1\over2}T_{bc}+T_{\beta\gamma})-\gamma G_{m}-\gamma^{2}b 
+\partial^2 c+\partial(c\xi\eta)
\label{eq:brs0}
\end{equation}
and
\begin{equation}
 T_{m}= T_{M}+T^{N=2}_{C},\quad\quad
 G_{m}= G_{M}+G^{+}_{C}+G^{-}_{C}.
\end{equation}
$Q_{BRST}$ is a nilpotent operator when a total central
charge is zero. In fact,
$T_{m}(z),G_{m}(z)$ satisfy 
$N=1$ superconformal algebra with $c_{m}=15$, while 
$T^{gh}(z),G^{gh}(z)$ satisfy $N=1$ superconformal algebra with $c_{gh}=-15$.
The Hilbert space ${\cal H}_{phys}$ is characterized by the BRST
cohomology;
${\cal H}_{phys}={\rm Ker}Q_{BRST}/{\rm Im}Q_{BRST}$.
Here $Q_{BRST}$ acts on the small Hilbert space
${\cal H}_{small}$ spanned by the ghost sector $(b,c,\beta,\gamma)$
and the matter part.
When we study the BRST cohomology, it is convenient to consider the
large Hilbert space ${\cal H}_{large}$ rather than ${\cal H}_{small}$.
The large Hilbert space is obtained from the 
small Hilbert space  by adding the zero mode $\xi_{0}$ of fermionic 
ghost $\xi$.
In the large Hilbert space, the physical state condition becomes
\begin{eqnarray}
 Q_{BRST}|\psi\rangle&=&0, \quad
|\psi\rangle\sim |\psi\rangle+Q_{BRST}|\Lambda\rangle,\nonumber\\
\eta_{0}|\psi\rangle&=&\eta_{0}|\Lambda\rangle=0,
\label{eq:phys1}
\end{eqnarray}
where 
$\eta_{0}$ is the zero-mode of $\eta$ which is conjugate to $\xi_{0}$.

It has been noticed in \cite{BeVa} that the BRST symmetry in 
superstring theory can be regarded as a part of topological 
$N=4$ superconformal symmetry.
The physical state conditions are translated into the chiral state
conditions in superconformal field theory.
In the hybrid formalism, this world-sheet symmetry comes from a part of
$\kappa$-symmetry\cite{Be,STVZ}.
The topological $N=4$ superconformal algebra is generated by the 
currents
\begin{eqnarray}
 J(z)&=&cb-\xi\eta, \quad
J^{++}(z)=c\eta,\quad
J^{--}(z)= \xi b, 
\nonumber\\
G^{+}(z)&=& J_{BRST}=
c(T_{m}+{1\over2}T_{bc}+T_{\beta\gamma})-\gamma G_{m}-\gamma^{2}b 
+\partial^{2}c +\partial(c\xi\eta), \nonumber\\
G^{-}(z)&=& b,\nonumber\\
\tilde{G}^{+}(z)&=&\eta, \nonumber\\
\tilde{G}^{-}(z)&=& \xi (T_{m}+T_{bc}+T_{\beta\gamma})+ e^{\phi}G_{m}b 
-\eta e^{2\phi} (\partial b) b -b c\partial\xi+\partial^{2} \xi,
\nonumber\\
T(z)&=& T_{m}+T_{bc}+T_{\beta\gamma}.
\label{eq:top}
\end{eqnarray}
The untwisted energy-momentum tensor $T^{N=2}=T-{1\over2}\partial
J$
has central charge $c=6$.
Now the physical state conditions (\ref{eq:phys1}) become
\begin{eqnarray}
G^{+}_{0}|\psi\rangle&=&0, \quad
|\psi\rangle\sim |\psi\rangle+G^{+}_{0}|\Lambda\rangle,\nonumber\\
\tilde{G}^{+}_{0}|\psi\rangle&=&\tilde{G}^{+}_{0}|\Lambda\rangle=0.
\label{eq:phys2}
\end{eqnarray}
Since the $\tilde{G}^{+}_{0}$-cohomology is trivial, {\it i.e.}
for a state $|\psi\rangle$ satisfying $\tilde{G}^{+}_{0}|\psi\rangle=0$
there is a state $|V\rangle\in {\cal H}_{large}$ such that
$|\psi\rangle=\tilde{G}^{+}_{0}|V\rangle$.
The physical state conditions  for $|V\rangle$
become
\begin{eqnarray}
 G^{+}_{0}\tilde{G}^{+}_{0}|V\rangle=0,\quad
|V\rangle\sim |V\rangle+
G^{+}_{0}|\Lambda\rangle+\tilde{G}^{+}_{0}|\tilde{\Lambda}\rangle.
\end{eqnarray}

We now investigate the space-time supersymmetry. 
In order to introduce space-time spinor, we bosonize the
worldsheet fermions $\psi^{+}=e^{i H_{1}}$, $\psi^{-}=e^{-i H_{1}}$,
where $H_{1}(z)H_{1}(w)\sim-\ln(z-w)$. 
The space-time supercharges are given by
$Q_{\alpha}^{\pm}=\int {dz\over2\pi i} q_{\alpha}^{\pm}(z)$ ($\alpha=1,2$)
where $q_{\alpha}^{\pm}(z)$ are the supercurrents
in the $-{1\over2}$ picture:
\begin{equation}
 q_{\alpha}^{\pm}=e^{-{\phi\over2}+{i\epsilon_\alpha H_{1}\over2}
\pm {i\over2}H_{C}}, 
\end{equation}
where $H_{C}$ is defined by $J_{C}=i\partial H_{C}$ and $\epsilon_1=1$, 
$\epsilon_2=-1$.
{}From the GSO projection, we may take $q_{1}^{\pm}(z)$ for mutually 
local operators.
The supersymmetry algebra is 
\begin{equation}
 \{ Q^{+}_{1}, Q^{-}_{1}\}=\int {dz\over 2\pi i}
e^{-\phi}\psi^{+}(z).
\end{equation}
Due to the picture number, supersymmery is not manifest
in the NSR formalism. 

In order to realize the space-time supersymmetry manifestly, we
change the picture number of $q^{+}_{1}$ from $-{1\over2}$ to
$+{1\over2}$ 
by applying the picture changing operator $Z=Q_{BRST}\xi$;
\begin{equation}
 Z=c\partial\xi-e^{\phi}G_{m}-\partial\eta e^{2\phi}b +\partial
(\eta e^{2\phi}b).
\end{equation}
Then $q^{+}_{1}$ becomes
\begin{equation}
 q_{1}^{+}
=\eta b e^{{3\phi\over2}+{i H_{1}\over2}+{i H_{C}\over2}}
+i\partial X^{+} e^{{\phi\over2}-{i H_{1}\over2}+{i H_{C}\over2}}
-G_{C}^{-}e^{{\phi\over2}+{i H_{1}\over2}+{i H_{C}\over2}}.
\end{equation}
The new space-time supercharges
\begin{equation}
Q^{\pm}_{1}=\int {dz\over 2\pi i} q_{1}^{\pm}(z)
\end{equation}
obey the anti-commutation relation
\begin{equation}
 \{ Q^{+}_{1}, Q^{-}_{1}\}=\int {dz\over 2\pi i}
i\partial X^{+}.
\label{eq:susy1}
\end{equation}
In this realization, certain linear combinations
of the fields $\phi$, $H_{1}$ and $H_{C}$ are important.
We define the new variables
\begin{equation}
 p_{1}=e^{-{1\over2}\phi+ {i\over 2}H_{1}- {i\over2} H_{C}},\quad
 p_{2}=\eta b e^{{3\over2}\phi+{i \over2}H_{1}+{i\over2}H_{C}},
\end{equation}
and their conjugate fields
\begin{equation}
\theta^{1}= e^{{1\over2}\phi-{i\over2}H_{1}+{i\over2}H_{C}}, \quad
 \theta^{2}=c\xi e^{-{3\over2}\phi-{i\over2}H_{1}-{i\over2}H_{C}},
\end{equation}
satisfying $p_{a}(z)\theta^{b}(w)\sim{\delta_{a}^{b}\over z-w}$.
Then the supercurrents become
\begin{eqnarray}
 q^{-}_{1}&=& p_{1}, \nonumber\\
 q^{+}_{1}&=&p_{2}+i\partial X^{+}\theta^{1}
-G_{C}^{-}e^{{\phi\over2}+{i H_{1}\over2}+{i H_{C}\over2}}.
\end{eqnarray}
The fermionic system $(p_{a},\theta^{a})$ has conformal weights $(1,0)$
and regarded as fundamental fields in the hybrid superstrings.
The fermions $\theta^{a}$ are
fermionic coordinates of target superspace.
Let us express them in the bosonized form $p_{a}=e^{i\phi_{a}}$
and $\theta^{a}=e^{-i\phi_{a}}$
($a=1,2$), 
where
\begin{eqnarray}
i \phi_{1}&=&  -{1\over2}\phi+{i\over2}H_{1}-{i\over2}H_{C},
\nonumber\\
i \phi_{2}&=& i\sigma+i\chi+{3\over2}\phi+{i\over2}H_{1}+{i\over2}H_{C}.
\end{eqnarray}
Since $(p_{a},\theta^{a})$ are not orthogonal to the $U(1)$ scalar field
$H_{C}$ and the ghost fields, we define the new $U(1)$ field
$\widehat{H}_{C}$ and the free bosons$(\rho, f)$.
\begin{eqnarray}
 i\widehat{H}_{C}&=&i H_{C}+4 (\phi+i \chi),
\nonumber\\
\rho&=&2\sqrt{2}\phi+{3\over\sqrt{2}}i\chi+{1\over\sqrt{2}}i H_{C}
+{1\over\sqrt{2}}i\sigma,
\nonumber\\
if&=&{1\over\sqrt{2}}(-\phi-i\chi+i H_{1}-i\sigma).
\end{eqnarray}
These are orthogonal to $(p_{a},\theta^{a})$.
Due to the change of the $U(1)$ current $J_{C}$, the generators of 
$N=2$ superconformal algebra are modified as 
\begin{eqnarray}
 \widehat{J}_{C}&=& i\partial \widehat{H}_{C},\quad
\widehat{T}^{N=2}_{C}=T^{N=2}_{C}+{1\over2} i\partial H_{C}\partial(\phi+i\chi)
+2 (\partial\phi+i\partial\chi)^2,
\nonumber\\
\widehat{G}^{\pm}_{C}&=&e^{\pm (\phi+i\chi)}G^{\pm}_{C}.
\end{eqnarray}
The hybrid superstrings are described by the free fields
$X^{\pm}$, $(p_{a},\theta^{a})$ ($a=1,2$), $\rho$, $f$ and 
$N=2$ SCFT $(\widehat{J}_{C}, \widehat{G}^{\pm}_{C},\widehat{T}^{N=2}_{C})$.
In terms of these new variables, the space-time supercurrents take the
form
\begin{equation}
 q^{-}_{1}= p_{1},
\quad
q^{+}_{1}= p_{2}+i\partial X^{+}\theta^{1}-e^{{1\over\sqrt{2}}(\rho+i f)}
\widehat{G}^{-}_{C}.
\label{eq:super1}
\end{equation}
The last term in $q^{+}_{1}$ is removed by the similarity transformation
${\cal O}\rightarrow 
({\cal O})^{R}\equiv e^{R}{\cal O} e^{-R}$ with the operator 
$R=\int {dz\over 2\pi i} c\widehat{G}^{-}_{C}$:
\begin{eqnarray}
 (q^{-}_{1})^{R}&=& p_{1},
\nonumber\\
(q^{+}_{1})^{R}&=& p_{2}+i\partial X^{+}\theta^{1}.
\label{eq:super2}
\end{eqnarray}
Further similarity transformation by the operator 
$U=\int {dz\over 2\pi i}{-i\over2}\theta^{1}\theta^{2}\partial X^{+}$
leads to the supercurrennts realized in a more symmetric way:
\begin{eqnarray}
 (q^{-}_{1})^{R+U}&=& p_{1}+{i\over2 }\partial X^{+}\theta^{2},
\nonumber\\
(q^{+}_{1})^{R+U}&=& p_{2}+{i\over2}\partial X^{+}\theta^{1}.
\label{eq:super3}
\end{eqnarray}
It is easy to see that the 
supercharges $Q^{\pm}_{1}=\int {dz\over 2\pi i}(q_{1}^{\pm})^{R+U}$ satisfy
the anticommutation relation (\ref{eq:susy1}).
Combining these with the anti-holomorphic part, we get $N=(2,2)$ 
supersymmetry
for type IIA or $N=(4,0)$ supersymmetry for type IIB.

We study the topological $N=4$ structure in the hybrid superstrings.
When we perform the similarity transformation by $R+U$, we may check
that the energy-momentum
tensor $T$ and $SU(2)$ current $J$, $J^{\pm\pm}$ are invariant.
In the currents $G^{\pm}$, the Calabi-Yau part and the flat space-time
part decouple after the transformation.
In terms of hybrid variables $(X^m,p^{a},\theta_{a},\rho,f,
\widehat{J}_{C}, \widehat{G}^{\pm}_{C},\widehat{T}^{N=2}_{C})$ 
the generators (\ref{eq:top})
are shown to be expressed 
as follows:
\begin{eqnarray}
(J)^{R+U}&=&-\sqrt{2}\partial\rho+i\partial\widehat{H}_{C}, \quad
(J^{\pm\pm})^{R+U}=e^{\pm (-\sqrt{2}\rho+i\widehat{H}_{C})},\nonumber\\
(T)^{R+U}&=& 
-\partial X^{+}\partial X^{-}
-p_{1}\partial\theta^{1}
-p_{2}\partial\theta^{2}-{1\over2}(\partial\rho)^{2}
-{1\over2}(\partial f)^{2}
-{1\over\sqrt{2}}\partial^{2}(\rho-i f)\nonumber\\
&&
+\widehat{T}^{N=2}_{C}+{1\over2}\partial \widehat{J}^{N=2}_{C},
\nonumber\\
(G^{+})^{R+U}&=&
e^{{1\over\sqrt{2}}(\rho+i f)} 
\left[
\left(
-i\partial X^{-}+{1\over2}(\theta^1\partial\theta^2-\partial\theta^1\theta^2)
\right)
\left(p_{1}-{i\over2}\theta^2\partial X^{+}\right)
-i\sqrt{2}\partial f \partial \theta^{2}
-{3\over4}\partial^2\theta^2
\right]\nonumber\\
&&+e^{{1\over\sqrt{2}}(\rho-i f)}(p_{1}-{i\over2}\theta^{2}\partial X^{+})
+\widehat{G}^{+}_{C}, \nonumber\\
(G^{-})^{R+U}&=& e^{-{i\over\sqrt{2}}(\rho+if)}
(p_{2}-{i\over2}\theta^1\partial X^{+})+\widehat{G}^{-}_{C}.
\label{eq:top2}
\end{eqnarray}
In the supercurrents $\tilde{G}^{\pm}$, the flat space-time part and
the Calabi-Yau part do not decouple. One finds that
\begin{eqnarray}
(\tilde{G}^{+})^{R+U}&=&
e^{-{1\over\sqrt{2}}(3\rho+i f)+i\widehat{H}_{C}}
(p_{2}-{i\over2}\theta^1\partial X^{+})
+e^{-\sqrt{2}\rho+i\widehat{H}_{C}}\widehat{G}^{-}_{C},\nonumber\\
(\tilde{G}^{-})^{R+U}&=&
e^{{1\over\sqrt{2}}(3\rho+i f)-i\widehat{H}_{C}} 
\left[
\left(
-i\partial X^{-}+{1\over2}(\theta^1\partial\theta^2-\partial\theta^1\theta^2)
\right)
\left(p_{1}-{i\over2}\theta^2\partial X^{+}\right)
-i\sqrt{2}\partial f \partial \theta^{2}
-{3\over4}\partial^2\theta^2
\right]\nonumber\\
&&+e^{{1\over\sqrt{2}}(3\rho-i f)-i\widehat{H}_{C}}
(p_{1}-{i\over2}\theta^{2}\partial X^{+})
+e^{\sqrt{2}\rho-i\widehat{H}_{C}}\widehat{G}^{+}_{C}.
\end{eqnarray}
In the following we shall use the variables after the similarity
transformations. We use ${\cal O}$ instead of ${\cal O}^{R+U}$ for
simplicity.

As in the case of four or six dimensional superstrings\cite{Be}, 
these currents
are written in terms of supercovariant derivatives.
Let $d_{a}$ ($a=1,2$) be supercovariant derivatives which anticommute
with $q^{\pm}_{1}(z)$
\begin{eqnarray}
 d_{1}&=& p_{1}-{i\over2}\theta^{2}\partial X^{+},\nonumber\\
 d_{2}&=& p_{2}-{i\over2}\theta^{1}\partial X^{+}.
\end{eqnarray}
We introduce also
\begin{eqnarray}
\pi^{+}&=&i\partial X^{+},\nonumber\\
 \pi^{-}&=& i\partial X^{-}+
{1\over2}(\partial \theta^{1} \theta^{2}-\theta^{1}\partial
\theta^{2}),\nonumber\\
\omega^{1}&=&\partial\theta^{1},\quad
\omega^{2}=\partial \theta^{2}.
\end{eqnarray}
They satisfy the following algebra:
\begin{eqnarray}
d_{1}(z)d_{2}(w)&&\sim {-\pi^{+}(w)\over (z-w)^2},\quad
\pi^{+}(z)\pi^{-}(w)\sim{1\over (z-w)^2},\nonumber\\
d_{1}(z)\pi^{-}(w)&&\sim {-\omega^{2}(w)\over z-w},\quad
d_{2}(z)\pi^{-}(w)\sim{-\omega^{1}(w)\over z-w},\nonumber\\
d_{1}(z)\omega^{1}(w)&&\sim {1\over (z-w)^2},\quad
d_{2}(z)\omega^{2}(w)\sim{1\over (z-w)^2}.
\end{eqnarray}
Then the generators of the  topological $N=4$ algebra become
\begin{eqnarray}
J&=&-\sqrt{2}\partial\rho+i\partial\widehat{H}_{C}, \quad
J^{\pm\pm}=e^{\pm (-\sqrt{2}\rho+i\widehat{H}_{C})},\nonumber\\
T&=& \pi^{+}\pi^{-}-d_{1}\partial\theta^{1}
-d_{2}\partial\theta^{2}
-{1\over2}(\partial\rho)^{2}
-{1\over2}(\partial f)^{2}
-{1\over\sqrt{2}}\partial^{2}(\rho-i f)\nonumber\\
&&
+\widehat{T}^{N=2}_{C}+{1\over2}\partial \widehat{J}_{C},
\nonumber\\
G^{+}&=&
e^{{1\over\sqrt{2}}(\rho+i f)} 
\left(-\pi^{-}d_{1}
-i\sqrt{2}\partial f \partial \theta^{2}
-{3\over4}\partial^2\theta^2
\right)
+e^{{1\over\sqrt{2}}(\rho-i f)}d_{1}
+\widehat{G}^{+}_{C}
, \nonumber\\
G^{-}&=& e^{-{1\over\sqrt{2}}(\rho+if)}d_{2}+\widehat{G}^{-}_{C},
\nonumber\\
 \tilde{G}^{+}&=&
e^{-{1\over\sqrt{2}}(3\rho+i f)+i\widehat{H}_{C}}
d_{2}
+e^{-\sqrt{2}\rho+i\widehat{H}_{C}}\widehat{G}^{-}_{C},\nonumber\\
\tilde{G}^{-}&=&
e^{{1\over\sqrt{2}}(3\rho+i f)-i\widehat{H}_{C}} 
\left(-\pi^{-}d_{1}
-i\sqrt{2}\partial f \partial \theta^{2}
-{3\over4}\partial^2\theta^2
\right)
+e^{{1\over\sqrt{2}}(3\rho-i f)-i\widehat{H}_{C}}d_{1}
+e^{\sqrt{2}\rho-i\widehat{H}_{C}}\widehat{G}^{+}_{C}. \nonumber\\
\label{eq:top3}
\end{eqnarray}
The space-time supersymmetry generators (\ref{eq:super3}) commute with 
$G^{+}_{0}$ and $\tilde{G}^{+}_{0}$.
The space-time supermultiplet of a observable 
belongs to the BRST cohomology. 
We note that $G^{+}$ can be decomposed in three mutually anticommuting
parts:
\begin{equation}
G^{+}=G^{+}_{(1)}+G^{+}_{(2)}+G^{+}_{C},
\label{eq:brs1}
\end{equation}
where $G^{+}_{(1)}=e^{{1\over\sqrt{2}}(\rho+i f)} 
\left(-(\pi^{-}d_{1})-i\sqrt{2}\partial f \partial \theta^{2}
-{3\over4}\partial^{2}\theta^2\right)$  and
$G^{+}_{(2)}= e^{{1\over\sqrt{2}}(\rho-i f)}d_{1}$.

Now we will study the hybrid superstrings on 
singular Calabi-Yau fourfolds.
A singular Calabi-Yau manifold $X^{8}$ is described by 
$R_{\varphi}\times S^{1}\times LG$\cite{GKP},
where $R_{\varphi}\times S^{1}$ denotes the $N=2$ supersymmetric 
Liouville theory with the central charge $c_{L}=3+3Q^2$.
$LG$ is an $N=2$ Laundau-Ginzburg model associated with 
$ADE$ type singularity and has the central charge $c_{LG}=12-c_{L}=9-3Q^2$.

The $N=2$ Liouville part 
is realized by two complex boson
$\Phi={1\over\sqrt{2}}(\varphi+i Y)$,
$\bar{\Phi}={1\over\sqrt{2}}(\varphi-i Y)$
and complex fermions
$\Psi={1\over\sqrt{2}}(\Psi^{1}+ i\Psi^{2})$,
$\bar{\Psi}={1\over\sqrt{2}}(\Psi^{1}- i\Psi^{2})$,
where $\varphi$ is a Liouville field with background charge $Q$ 
and $Y$ is a free boson compactified on $S^1$.
The OPEs are given by
$\varphi(z)\varphi(w)\sim -\ln(z-w)$, $Y(z)Y(w)\sim -\ln (z-w)$
and $\Psi^{a}(z)\Psi^{b}(w)\sim \delta^{ab}/(z-w)$ $(a,b=1,2)$.
The worldsheet $N=2$ algebra with the central charge $c_{L}=3+3Q^2$ 
can be obtained from the generators:
\begin{eqnarray}
 T_{L}&=&-\partial\Phi\partial\bar{\Phi}
-{Q\over2\sqrt{2}}\partial^2(\Phi+\bar{\Phi}) 
-{1\over2} \bar{\Psi}\partial\Psi
-{1\over2} \Psi\partial\bar{\Psi},\nonumber\\
G^{+}_{L}&=&i\Psi\partial \bar{\Phi}
+{i Q\over\sqrt{2}}\partial \Psi,
\quad
G^{-}_{L}=i\bar{\Psi}\partial \Phi
+{i Q\over\sqrt{2}}\partial \bar{\Psi},
\nonumber\\
J_{L}&=&\Psi\bar{\Psi}+iQ\partial Y.
\end{eqnarray}
For the Landau-Ginzburg part, we write the generators of $N=2$ algebra as
$J_{LG},G^{\pm}_{LG}, T^{N=2}_{LG}$
Then we have
\begin{eqnarray}
\widehat{J}_{C}&=& J_{L}+J_{LG},\quad 
\widehat{T}^{N=2}_{C}=T^{N=2}_{L}+T^{N=2}_{LG}\nonumber\\
\widehat{G}^{\pm}_{C}&=& G^{\pm}_{L}+G^{\pm}_{LG}.
\end{eqnarray}
In \cite{GKP}, it is argued that the superstrings on singular
Calabi-Yau manifolds
correspond to the space-time $N=2$  superconformal
field theory with central charge $c=6kp$ in the decoupling limit.
Here $k$ is related to the Liouville background charge $Q$ by the 
formula $k=2/Q^2$ and $p$ is an integer.

We expect that the space-time $N=2$ supersymmetry (\ref{eq:super3}) is 
enhanced to $N=2$
superconformal symmetry generated by
${\cal L}_{n}$, ${\cal I}_{n}$ $(n\in {\bf Z})$ and
${\cal G}^{\pm}_{r}$ $(r\in {\bf Z}+{1\over2})$.
Their algebra is
\begin{eqnarray}
\left[{\cal L}_n,{\cal L}_m\right]&=&
(n-m){\cal L}_{n+m}+\frac{c}{12}n(n^2-1)\delta_{n+m,0},\nonumber\\
\left[{\cal L}_n,{\cal G}^{\pm}_r\right]&=&
\left(\frac{n}{2}-r\right){\cal G}^{\pm}_{n+r},\nonumber\\
\left[{\cal L}_n,{\cal I}_m\right]&=&
-m{\cal I}_{n+m},\nonumber\\
\left\{{\cal G}^+_r,{\cal G}^-_s\right\}&=&
{\cal L}_{r+s}+\frac{1}{2}(r-s){\cal I}_{r+s}
+\frac{c}{6}\left(r^2-\frac{1}{4}\right)\delta_{r+s,0},\nonumber\\
\left[{\cal I}_n,{\cal G}^{\pm}_r\right]&=&
\pm{\cal G}^{\pm}_{n+r},\nonumber\\
\left[{\cal I}_n,{\cal I}_m\right]&=&
\frac{c}{3}n\delta_{n+m,0}.
\end{eqnarray}
${\cal G}^{\pm}_{-{1\over2}}$ generate the $N=2$ supersymmetry and 
${\cal L}_{-1}$ corresponds to the translation:
${\cal G}^{\pm}_{-{1\over2}}=Q^{\pm}_{1}$, ${\cal L}_{-1}=
\int {dz\over 2\pi i}i\partial X^{+}$.

We discuss a finite dimensional subalgebra of $N=2$
superconformal symmetry.
In terms of zero-modes of superspace coordinates $(X^{-},\theta^a)$, the 
$N=2$ space-time superconformal generators 
${\cal L}_{0},{\cal L}_{\pm1},{\cal G}_{\pm{1\over2}},
\bar{\cal G}_{\pm{1\over2}},{\cal I}_{0}$ are given by
\begin{eqnarray}
 {\cal L}_{-1}&=&-i{\partial\over\partial X^{-}},\nonumber\\
{\cal L}_{0}&=&-X^{-}{\partial\over\partial X^{-}}
-{1\over2}\left(\theta^{1}{\partial\over\partial \theta^{1}}+
\theta^{2}{\partial\over\partial \theta^{2}}\right),\nonumber\\
{\cal L}_{1}&=&i(X^{-})^2{\partial\over\partial X^{-}}
-iX^{-}\left(\theta^{1}{\partial\over\partial \theta^{1}}+
\theta^{2}{\partial\over\partial \theta^{2}}\right),\nonumber\\
{\cal G}_{{1\over2}}^{+}&=&
-i X^{-}\left(
{\partial\over\partial \theta^{2}}
-{i\over2}\theta^{1}{\partial\over\partial X^{-}}\right)
-{1\over 2}\theta^{1}\theta^{2}{\partial\over\partial \theta^{2}},
\nonumber\\
{\cal G}_{-{1\over2}}^{+}&=&
{\partial\over\partial \theta^{2}}
-{i\over2}\theta^{1}{\partial\over\partial X^{-}},\nonumber\\
{\cal G}_{{1\over2}}^{-}&=&
-i X^{-}\left(
{\partial\over\partial \theta^{1}}
-{i\over2}\theta^{2}{\partial\over\partial X^{-}}\right)
-{1\over 2}\theta^{1}\theta^{2}{\partial\over\partial \theta^{1}},
\nonumber\\
{\cal G}_{-{1\over2}}^{-}&=&
{\partial\over\partial \theta^{1}}
-{i\over2}\theta^{2}{\partial\over\partial X^{-}},
\nonumber\\
{\cal I}_{0}&=&
\theta^{1}{\partial\over\partial \theta^{1}}-
\theta^{2}{\partial\over\partial \theta^{2}}.
\end{eqnarray}
To realize the infinite dimensional algebra, it is convenient to
introduce $(\beta,\gamma)$ system instead of $(X^{+},X^{-})$ by
\begin{equation}
 \beta=i\partial X^{+},\quad \gamma=-i X^{-}.
\end{equation}
which has the OPE $\beta(z)\gamma(w)\sim (-1)/(z-w)$.
The space-time $N=2$ superconformal algebra is now realized in terms of
hybrid variables $(\beta,\gamma)$, $(p_{a},\theta^{a})$ and $(\rho,f)$.
In addition, ${\cal I}_{0}$ corresponds to the ${\cal R}$-charge of the 
theory and is characterized by the $U(1)$ current $i\partial Y$.
In the $AdS_{3}$ case\cite{GiKuSe}, we need the Liouville field to 
construct the space-time Virasoro algebra.
In the present case, $\varphi$ plays a similar role.
These field contents lead to the free field realization of currents of 
affine Lie superalgebra $sl(2|1)^{(1)}$ at level $k$\cite{It0}:
\begin{eqnarray}
\hat{J}^{++}&=&
\gamma^2\beta-\gamma\left(
\theta^1p_1+\theta^2p_2-\partial(R-a\Phi)+\partial(\bar R+a\bar\Phi)
\right)\nonumber\\
&&
+(k-1)\frac{1}{2}(\theta^1\partial\theta^2-\partial\theta^1\theta^2)
-\frac{1}{2}\theta^1\theta^2\left(
\partial(R-a\Phi)+\partial(\bar R+a\bar\Phi)\right)
-k\partial\gamma
,\nonumber\\
\hat{J}^3&=&
-\gamma\beta+\frac{1}{2}\left(
\theta^1p_1+\theta^2p_2-\partial(R-a\Phi)+\partial(\bar R+a\bar\Phi)
\right),\nonumber\\
\hat{J}^{--}&=&\beta,\nonumber\\
\hat{I}&=&\frac{1}{2}\left(
\theta^1p_1-\theta^2p_2-\partial(R-a\Phi)-\partial(\bar R+a\bar\Phi)
\right),\nonumber\\
\hat{j}^{+(+)}&=&
\gamma(p_2+\frac{1}{2}\theta^1\beta)-(k-\frac{1}{2})\partial\theta^1
-\theta^1\partial(\bar R+a\bar\Phi)-\frac{1}{2}\theta^1\theta^2p_2,
\nonumber\\
\hat{j}^{+(-)}&=&
\gamma(p_1+\frac{1}{2}\theta^2\beta)-(k-\frac{1}{2})\partial\theta^2
+\theta^2\partial(R-a\Phi)+\frac{1}{2}\theta^1\theta^2p_1
,\nonumber\\
\hat{j}^{-(+)}&=&
p_2+\frac{1}{2}\theta^1\beta,\nonumber\\
\hat{j}^{-(-)}&=&
p_1+\frac{1}{2}\theta^2\beta.
\end{eqnarray}
where $a=\sqrt{k}=\frac{\sqrt{2}}{Q}$ and
\begin{equation}
R=\frac{1}{\sqrt{2}}(\rho+if),\qquad
\bar R=\frac{1}{\sqrt{2}}(\rho-if).
\end{equation}
{}From the affine Lie superalgebra,
one may construct the space-time $N=2$  superconformal algebra as
discussed in the $AdS_{3}$ case\cite{It}:
\begin{eqnarray}
{\cal L}_n&=&\oint{dz\over 2\pi i}\left(\frac{1}{2}n(n+1)\gamma^{n-1}\left(
\hat{J}^{++}+{k\over2}\partial(\theta^{1}\theta^{2})\right) 
+(n^2-1)\gamma^n \hat{J}^3+\frac{1}{2}n(n-1)\gamma^{n+1} \hat{J}^{--}\right),
\nonumber\\
{\cal G}^+_r&=&\oint{dz\over 2\pi i}\left(
(r+\frac{1}{2})\gamma^{r-\frac{1}{2}}( \hat{j}^{+(+)}+k\partial\theta^1)
-(r-\frac{1}{2})\gamma^{r+\frac{1}{2}} \hat{j}^{-(+)}\right),
\nonumber\\
{\cal G}^-_r&=&\oint{dz\over 2\pi i}\left(
(r+\frac{1}{2})\gamma^{r-\frac{1}{2}} \hat{j}^{+(-)}
-(r-\frac{1}{2})\gamma^{r+\frac{1}{2}}\hat{j}^{-(-)}
-(r^2-{1\over4})\gamma^{r-{3\over2}}{k-{1\over2}\over2}\theta^{1}\theta^{2}
\partial \theta^2
\right),
\nonumber\\
{\cal I}_n&=&\oint{dz\over 2\pi i}\left(
2\gamma^n\hat I-n\gamma^{n-1}\left(
(k-{1\over2})\theta^1\partial\theta^2
-{1\over2}\partial\theta^1\theta^2-{1\over2}
\theta^1\theta^2(\partial(R-a\Phi)-\partial(\bar{R}+a\bar{\Phi})\right)\right).
\nonumber\\
\label{eq:gen1}
\end{eqnarray}
These generators satisfy $N=2$ space-time superconformal algebra with
the central charge $c=6kp$, where
\begin{equation}
 p=\int {d z\over 2\pi i}{\partial \gamma\over \gamma}. 
\end{equation}
We must, however, examine consistency of this space-time $N=2$
 superconformal symmetry with the topological
$N=4$ structure of the theory. 
Since the physical symmetry
have to (anti-)commute with the BRST charge, the space-time
charges (\ref{eq:gen1}) must (anti-) commute with operators
$G^{+}_{0}=\int {dz\over 2\pi i}G^{+}(z)$ and
$\tilde{G}^{+}_{0}=\int {dz\over 2\pi i}\tilde{G}^{+}(z)$.
For $sl(2|1)$ currents, however,
it is shown that the currents $\hat{J}^{++}$ and $\hat{j}^{+(+)}$ are not
$G^{+}_{0}$-invariant.
Moreover, $\hat{J}^{++}$ and $\hat{j}^{+(-)}$ are not
$\tilde{G}^{+}_{0}$-invariant.
This implies that the original model proposed in \cite{GKP}
does not have the space-time superconformal symmetry
in their physical spectrum.
Here we modify the model in order to allow the space-time
superconformal symmetry.

We will solve the above problem by modifying both the worldsheet
topological currents and the space-time currents.
After some calculations, we are able to find the following solution.
Firstly, the $sl(2|1)$ currents
are modified as
\begin{equation}
 \hat{J}^{(new)++}=\hat{J}^{++}+aie^{-R}\theta^{1}\Psi,\quad
 \hat{j}^{(new)+(-)}=\hat{j}^{+(-)} -aie^{-R}\Psi,
\label{eq:gen2}
\end{equation}
and other currents are invariant.
The space-time $N=2$ superconformal generators are obtained by replacing
the affine currents to the new ones in (\ref{eq:gen1}). But
${\cal G}^{-}_{r}$ and ${\cal I}_{n}$
receive further corrections:
\begin{eqnarray}
 {\cal G}^{new -}_{r}&=& {\cal G}^{old -}_{r}
-\int {dz\over 2\pi i}\left((r^2-{1\over4})\gamma^{r-{3\over2}}{ai\over2}
\theta^1\theta^2 e^{-R}\Psi\right),\nonumber\\
{\cal I}^{new}_{n}&=&{\cal I}^{old}_{n}
-\int {dz\over 2\pi i} a i n\gamma^{n-1} e^{-R}\theta^1\Psi.
\label{eq:gen3}
\end{eqnarray}
The BRST currents which is consistent with these space-time currents
take the form
\begin{eqnarray}
G^{new +}&=&G^{+}_{(1)}+\Delta G^{+}_{(1)}+G^{+}_{L}+G^{+}_{LG}, \nonumber\\
\widetilde G^{new +}&=& \widetilde{G}^{old +}+\Delta \widetilde{G}^{+},
\label{eq:brs2}
\end{eqnarray}
where
\begin{eqnarray}
 \Delta G^{+}_{(1)}&=&e^{R}\left\{(\partial R+a\partial\bar{\Phi})
\partial\theta^2+\partial^2\theta^2\right\},  \nonumber\\
 \Delta \widetilde{G}^{+}&=& -a i e^{-\sqrt{2}\rho+i \widehat{H}_{C}}
(\bar{\Psi}\partial R+\partial \bar{\Psi}).
\end{eqnarray}
In $G^{+}(z)$, the term $G^{+}_{(2)}$ is eliminated in order to keep
the OPE with $G^{-}(z)$.
The energy-momentum tensor $T$ and $SU(2)$ currents $J^{\pm\pm},J^{3}$ 
do not change.
The supercurrents $G^{-}$ and $\tilde{G}^{-}$ are modified due to the 
change of $G^{+}$ and $\widetilde{G}^{+}$.
To summarize, the worldsheet $N=4$ currents are
\begin{eqnarray}
T
&=&-\beta\partial\gamma-p_1\partial\theta^1
-p_2\partial\theta^2-\partial\bar R\partial R
-\partial^2\bar R \nonumber\\
&&\hspace{10mm}
-\partial\bar\Phi\partial\Phi
-\frac{Q}{\sqrt{2}}\partial^2\bar\Phi
-\bar\Psi\partial\Psi+T_{LG}+\frac{1}{2}\partial J_{LG},\nonumber\\
G^+&=&
e^R\left(
-\pi^-d_1+\partial(\bar R+a\bar\Phi)\partial\theta^2
+\frac{1}{4}\partial^2\theta^2\right)
+i\left(\Psi
\partial\bar\Phi+\frac{1}{a}\partial\Psi\right)+G^+_{LG},
\nonumber\\
G^-&=&
e^{-R}d_2+i\left(\bar\Psi\partial(\Phi-aR)
-\left(a-\frac{1}{a}\right)\partial\bar\Psi\right)+G^-_{LG},
\nonumber\\
\widetilde G^+&=&
e^{-2R-\bar R+i H_{L}+i H_{LG}}d_2
+e^{-R-\bar R+i H_{L}+i H_{LG}}\left(\bar{\Psi}\partial(\Phi-aR)
+\left(a-\frac{1}{a}\right)\partial\bar{\Psi}\right)\nonumber\\
&&\hspace{20mm}
-e^{-R-\bar R+i H_{L}+i H_{LG}} G^{-}_{LG}, \nonumber\\
\widetilde G^-&=&
e^{2R+\bar R-i H_{L}-i H_{LG}}(-\pi^-d_1
+\partial(\bar R+a\bar\Phi)\partial\theta^2
+\frac{1}{4}\partial^2\theta^2)\nonumber\\
&&\hspace{10mm}
+e^{R+\bar R-i H_{L}-i H_{LG}}
\left(\Psi\partial\bar\Phi+\frac{1}{a}\partial\Psi\right)
-e^{R+\bar R-i H_{L}- i H_{LG}} G^{+}_{LG},\nonumber\\
J&=&-\partial(R+\bar R)+i \partial H_{L}+i\partial H_{LG},\nonumber\\
J^{++}&=&e^{-R-\bar R+i H_{L}+ i H_{LG}},\nonumber\\
J^{--}&=&e^{R+\bar R-i H_{L}-i H_{LG}},
\end{eqnarray}
where $J_{L}=i\partial H_{L}$ and $J_{LG}=i\partial H_{LG}$.
New $N=2$ superconformal generators (\ref{eq:gen1}), (\ref{eq:gen2}) 
and (\ref{eq:gen3}) are  consistent
with this $N=4$ topological structure. The new model allows
the space-time superconformal symmetry in the physical spectrum.
Based on these currents, we may study the BRST cohomology since
it is not clear that the new BRST current reproduces the consistent
physical spectrum with that in \cite{GKP}.
In the following paper\cite{ItKu}, 
we will investigate the BRST cohomology and
physical spectrum in detail.

In this paper, we have constructed the space-time $N=2$
superconformal algebra in the superstrings compactified
on singular Calabi-Yau fourfolds. We find that the original
BRST operator (\ref{eq:brso}) does not (anti-)commute with the 
superconformal
generators. The space-time supersymmetry of the original model
\cite{GKP}, therefore, does not extend to the superconformal.
We construct a new topological $N=4$ superconformal algebra
including the new BRST operator which commute with
the space-time superconformal generators. These generators
nontrivialy mix the two-dimensional and Liouville parts
as expected.

It is an interesting problem to generalize the present approach to the
case of $4$ or $6$ dimensional hybrid superstrings on a singular
Calabi-Yau three or two-fold, respectively. 
We would be able to study higher dimensional
superconformal field theory in terms of hybrid superstring formalism.
It is also interesting to apply the present approach to the case of 
the superstrings with RR backgrounds \cite{BeVaWi}
and the theory compactified on
$G_{2}$ and $Spin(7)$ holonomy manifolds\cite{ShVa}. 
These subjects will be discussed elsewhere.

\vspace{1cm}
{\bf Acknowledgments}:
We would like to thank the organizers of the Summer Institute 2001 at 
Yamanashi, where a part of this work was done.
The research of KI and HK is supported in part by the Grant-in-Aid from
the Ministry of Education, Science, Sports and Culture of Japan,
Priority Area 707 ``Supersymmetry and Unified Theory of Elementary
Particles'' and Priority Area No.13135213, respectively. HK is also
supported in part by the Grant-in-Aid for Scientific Research No.11640276
from Japan Society for Promotion of Science.

\end{document}